\documentclass{appolb}
\usepackage{epsfig}
\usepackage{dcolumn}
\usepackage{bm}
\usepackage{hyperref} 
\usepackage{multirow} 
\usepackage{multicol}
\newcommand{\Lambdabar}{\ensuremath{\overline{\Lambda}}}
\newcommand{\Omegabar}{\ensuremath{\overline{\Omega}}}
\newcommand{\MeV}{\ensuremath{\,{\rm MeV}}}
\newcommand{\GeV}{\ensuremath{\,{\rm GeV}}}
\newcommand{\req}[1]{Eq.\,(\ref{eq:#1})}
\newcommand{\labeq}[1]{\label{eq:#1}}
\long\def\symbolfootnotemark[#1]{\begingroup%
\def\thefootnote{\fnsymbol{footnote}}\footnotemark[#1]\endgroup} 
\begin{document}
\title{%
Strangeness Production in\\ Au--Au collisions at $\sqrt{s_{NN}}=62.4 \GeV$}
\author{%
Michal Petr{\' a}{\v n$^{1,2}$,
Jean Letessier$^{1,3}$,\\
Vojt{\v e}ch Petr{\' a}{\v c}ek$^2$,
Jan Rafelski$^1$}%
\address{%
$^1$Department of Physics, University of Arizona, Tucson, Arizona 85721\\
$^2$Czech Technical University in Prague,\\Faculty of Nuclear Sciences and Physical Engineering\\
$^3$Laboratoire de Physique Th{\' e}orique et Hautes Energies,\\
Universit{\' e} Paris 6, Paris 75005, France
}
}
\maketitle

\begin{abstract}
We obtain strangeness production as function of centrality in a statistical hadronization model analysis of all experimental hadron production data in  Au--Au collisions at $\sqrt{s_{NN}}=62.4\GeV$. Our analysis describes successfully the yield of strange and multi-strange hadrons recently published. We explore condition of hadronization  as a function of centrality and find universality for the case of chemical non-equilibrium in the hadron phase space corresponding to   quark--gluon plasma (QGP) in chemical equilibrium.
\end{abstract}
  
\section{Experimental Data}
The statistical hadronization model (SHM) is known to successfully describe hadron yields in a variety of experiments at different energies~\cite{Torrieri:2004zz,Becattini:2010sk}. Recently, new results on the yields of strange and multi-strange hadrons at $\sqrt{s_{NN}}=62.4\GeV$ have been published~\cite{Aggarwal:2010ig}. This data together with previously published non-strange particle yields~\cite{:2008ez} and $\phi$ meson measurement~\cite{:2008fd} provides a new opportunity to reevaluate hadronization conditions in $\sqrt{s_{NN}}=62.4 \GeV$ collisions. 

Heavier multistrange  particles are more rarely produced and therefore to assure that statistical errors are small, they are often presented in wider centrality bins than non-strange particles. In order to study all particle yields as a function of centrality, we have to account for the different centrality binning. We associate each bin with average number of participants $N_{\rm part}$. Then, we fit hadron yield as a function of $N_{\rm part}$ using a power law,
\begin{equation}
\labeq{interpol}
f(N_{\rm part}) = a \cdot N_{\rm part}^b + c,
\end{equation}
where $a,b$ and $c$ are free parameters fitted and shown in table \ref{tab:params}. We see as expected   that particles are always more numerous than their respective anti-particles.
However for the $\Omega$, we had only three data points and three free parameters of the interpolating function from \req{interpol}.  Therefore, we fixed  $c=0$ and fitted the other two. This fit, with on degree of freedom, gave us a qualitatively similar function as for the other particles. 

\begin{table}[h!]
\caption{\label{tab:params} Particle yields power fit parameters (as defined by \req{interpol}) used to describe particle yields as function of centrality.
For $\Omega$, c=0 is assumed} 
\begin{center}
\begin{tabular}{l|c|c|c}
	&	a	&	b	&	c	\\ \hline
$\pi^-$		&	$3.756\times 10^{-1}$	&	1.098	&	$-2.330\times 10^{-2}$	\\
$\pi^+$		&	$3.788\times 10^{-1}$	&	1.094	&	$-5.768\times 10^{-2}$	\\
$K^+$		&	$4.769\times 10^{-2}$	&	1.141	&	$-2.002\times 10^{-1}$	\\
$K^-$		&	$4.188\times 10^{-2}$	&	1.138	&	$-1.165\times 10^{-1}$	\\
$K_s^0$		&	$5.068\times 10^{-2}$	&	1.074	&	$-2.341\times 10^{-1}$	\\
$p^+$		&	$3.972\times 10^{-2}$	&	1.125	&	$-9.963\times 10^{-2}$	\\
$p^-$		&	$3.404\times 10^{-2}$	&	1.024	&	$-8.573\times 10^{-2}$	\\
$\phi^0$	&	$4.162\times 10^{-3}$	&	1.203	&	$-1.311\times 10^{-3}$	\\
$\Lambda$	&	$1.772\times 10^{-2}$	&	1.154	&	$-1.025\times 10^{-2}$	\\
$\Lambdabar$	&	$1.045\times 10^{-2}$	&	1.125	&	$\phantom{-}1.764\times 10^{-2}$	\\
$\Xi^-$	&	$1.434\times 10^{-3}$	&	1.197	&	$-1.415\times 10^{-2}$	\\
$\overline{\Xi}^+$&	$7.693\times 10^{-4}$	&	1.231	&	$-1.765\times 10^{-3}$	\\
$\Omega$	&	$1.266\times 10^{-5}$	&	1.720	&	$0.$	\\
$\Omegabar$	&	$3.985\times 10^{-6}$	&	1.879	&	$0.$	\\
\end{tabular}
\end{center}
\end{table}

\section{SHM fit including multistrange hadrons}

We performed a cross check of SHARE~\cite{Torrieri:2004zz}, the thermal model implementation we used, with three other codes currently in use by other groups using the compiled data in~\cite{Becattini:2010sk}. We tested our model on the data set from Au--Au collision at $\sqrt{s_{NN}} = 200\GeV$ from STAR and obtained both yields and thermal parameters within 5\% from the values in~\cite{Becattini:2010sk}.

Rather than hadron ratios studied in~\cite{Aggarwal:2010ig}, we decided to fit hadron yields, which require a common volume $dV/dy$ of the source. We complemented the data set with identified hadrons ($\pi^\pm, K^\pm, p^\pm$) from~\cite{:2008ez} and $\phi$ yield from~\cite{:2008fd}. Furthermore, we require the conservation of charge per baryon $(Q-\overline{Q})/(B-\overline{B}) = 0.39  \pm  0.01$ and strangeness $(s-\overline{s})/(s+\overline{s}) = 0 \pm 0.05$ with a few percent error to account for the detector acceptance and efficiency.

\begin{figure}[ht]
\begin{minipage}[t]{0.5\linewidth}
\includegraphics[width=3.3in,angle=-90]{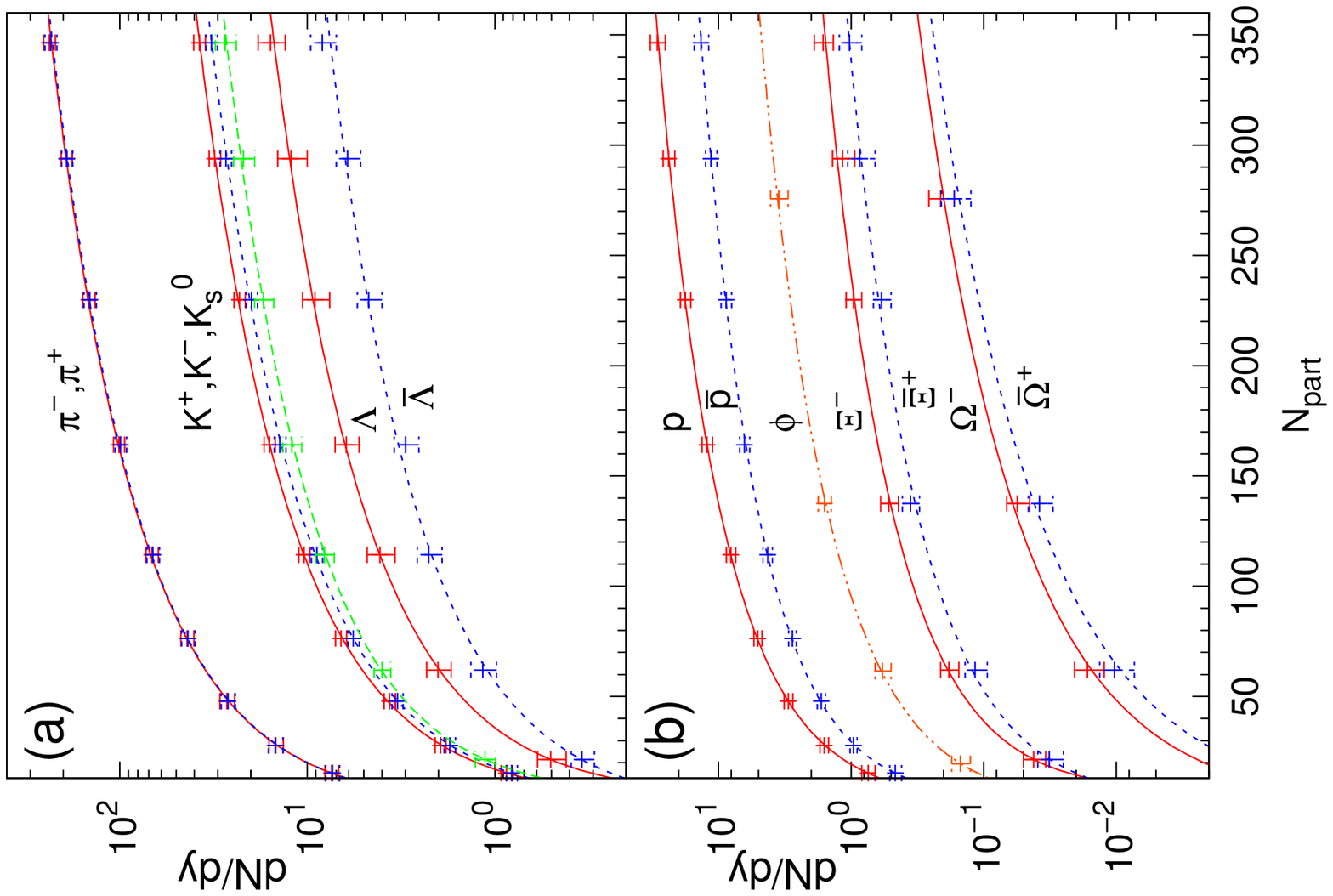}
\end{minipage}\hspace*{-0.3cm}
\begin{minipage}[t]{0.5\linewidth}
\includegraphics[width=3.3in,angle=-90]{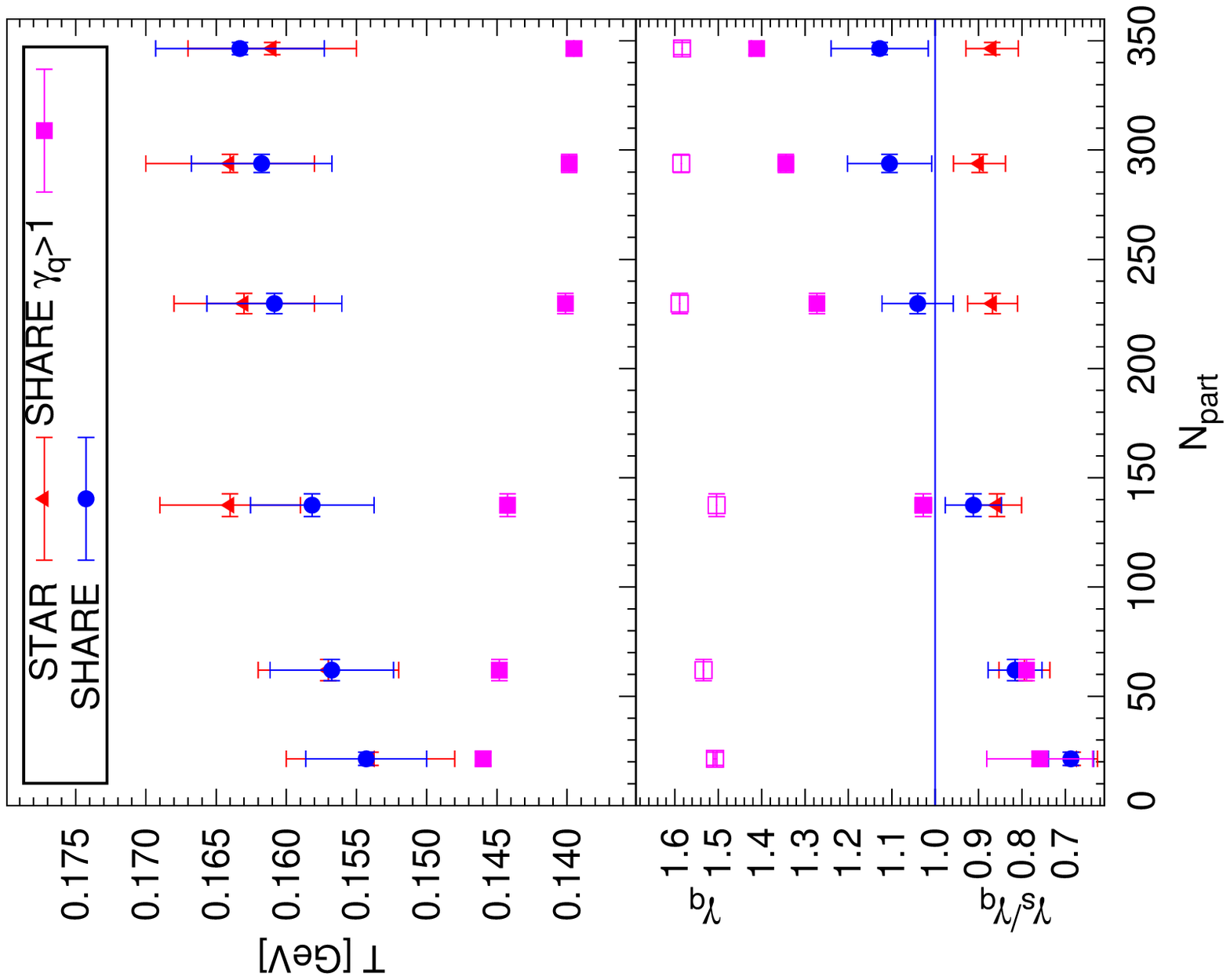}
\end{minipage}
\caption{\label{fig:yields}{\bf Left panel:} Particle yields as a function of number of participants with respective interpolating functions. 
{\bf Right panel:} \label{fig:tempgams} Top panel depicts chemical freeze-out temperature $T$ as a function of the number of participants $N_{part}$. Red triangles are a copy of the results in~\cite{Aggarwal:2010ig}, blue circles describe our fit of the same data set in semi-equilibrium model with $\gamma_q\equiv 1, \gamma_s\ne 1$ and purple squares are non-equilibrium model $\gamma_i\neq 1, i=q,s$. Bottom panel shows the strangeness phase space occupancy compared to the light quark phase space occupancy ratio $\gamma_s/\gamma_q$ for different models. Open purple squares show the values of $\gamma_q$ for the non-equilibrium model.}
\end{figure}

As was also pointed out in~\cite{Aggarwal:2010ig}, the experimental particle yields are subject to an important inconsistency: for all centralities, the yield  of $K_s^0$ is smaller than both $K^+$ and $K^-$ (lowest of three corresponding lines in the left panel of the figure \ref{fig:tempgams}). In general, at finite baryon density, we expect K$^+<$K$^0<$K$^-$. Interestingly the missing $K_s^0$ are  associated with yields of $\Lambda,\overline{\Lambda}$   in that these three particle types had the largest contributions to the total $\chi^2$ of the fit. The choice to include or exclude them from the fit did not change the values of the SHM parameters we obtained. 

It is interesting to note that the three neutral particle types have the same V-type charged particle decay and that there is an overlap in kinematic ID of the decay channels. For this reason we fitted instead the yield of $\Lambda+\overline{\Lambda}+2\cdot K_s^0$ which does not have this ambiguity. Considering the smallness of the required shift from more numerous   $K_s^0$ to  $\Lambda,\ \overline{\Lambda}$  we also fitted the ratio $\overline{\Lambda}/\Lambda$. These two data points  replace the three raw yields and proved to be consistent with all other particle yields. Assuming that other particles fix the source volume, this procedure incurs minimal loss of data information.

We expected that our SHM fit with $\gamma_q=1$ should reproduce results of~\cite{Aggarwal:2010ig}. And indeed, we obtain compatible chemical freeze-out temperature $T$ (top right panel of figure~\ref{fig:tempgams}). However,  we find that the strangeness phase space is strongly overpopulated, which is reflected by $\gamma_s > 1$ (bottom right panel of figure~\ref{fig:tempgams}) contrary to results published by STAR. After we found this discrepancy, we learned 
from the authors that in \cite{Aggarwal:2010ig} $\gamma_s < 1$ has been forced. This of course completely invalidates the conclusions of this work. This also resolves the myth that the yields of $\phi\propto \gamma_s^2$ are inconsistent with SHM.  

As a next step, we considered the chemical non-equilibrium with $\gamma_i \neq 1, i=q,s$. This approach allows that aside of the strangeness phase space also the light  $q=u,\ d$  quark phase space when observed on hadron side to be out of equilibrium. To preserve the number of degrees of freedom, we fitted the pressure $P\simeq 82 \MeV/{\rm fm}^3$, which is the value of critical hadronization pressure found in SHM fits to other  experiments~\cite{Rafelski:2009jr}. 

The chemical non-equilibrium SHM fit results in an even more pronounced strangeness yield enhancement, $\gamma_s \to  1.5$ at central rapidity,   an over-population of the light quark phase space  $\gamma_q \simeq 1.6$,  and a lower freeze-out temperature $T\simeq 142 \MeV$. Another important feature of this approach is that the parameter errors in the fit decreased dramatically compared to the semi-equilibrium model. In the right bottom panel of figure~\ref{fig:tempgams},  the centrality dependence of these parameters is shown and the errors on $\gamma_q$ and $\gamma_s$ 
for chemical non-equilibrium are within the size of the symbols. We note that the 
SHM fits at all centralities have  confidence level above 60\%.

\begin{figure}[h!t]
\begin{minipage}[t]{0.5\linewidth}
\includegraphics[width=3.0in,angle=-90]{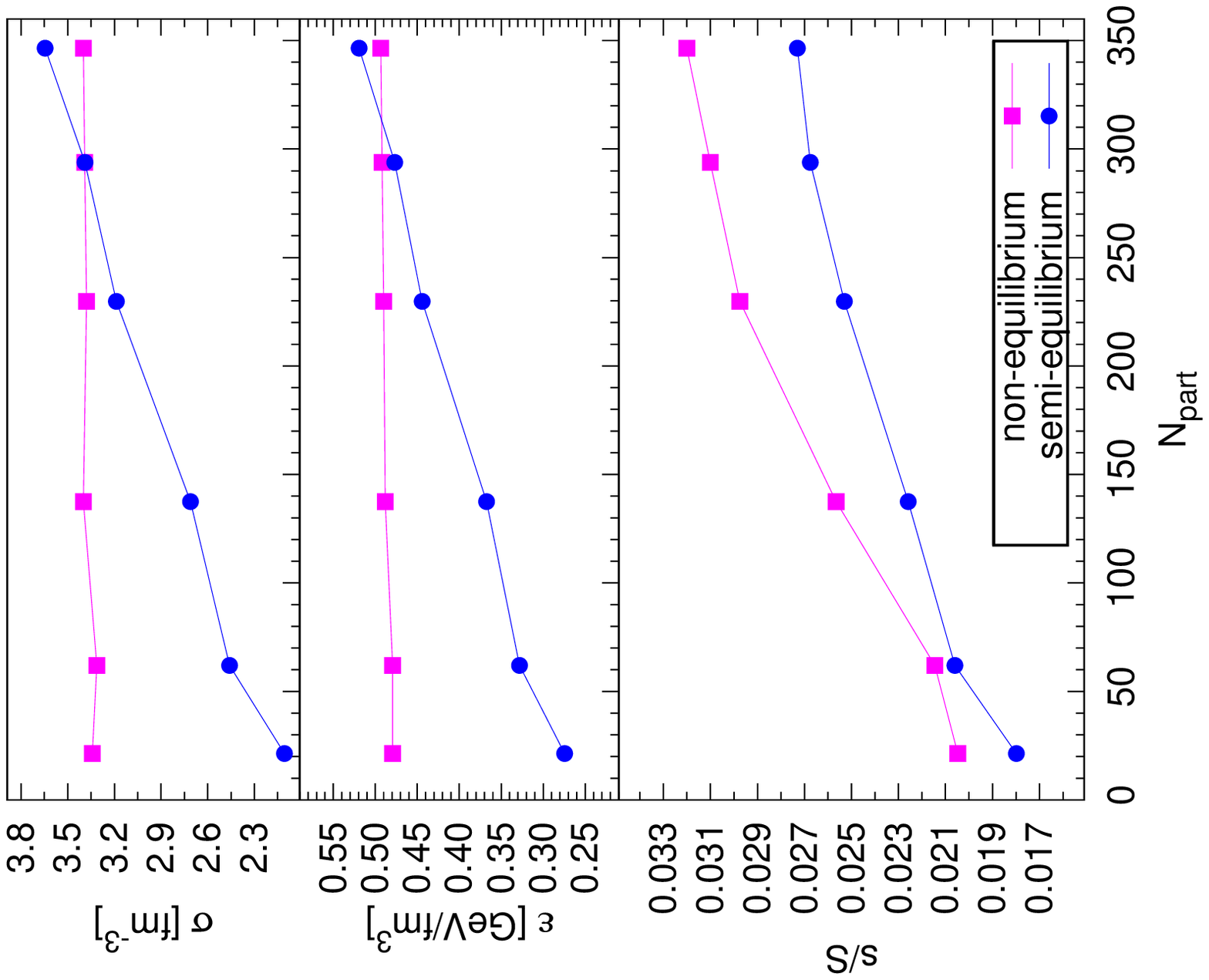}
\end{minipage}\hspace*{-0.3cm}
\begin{minipage}[t]{0.5\linewidth}
\includegraphics[width=3.0in,angle=-90]{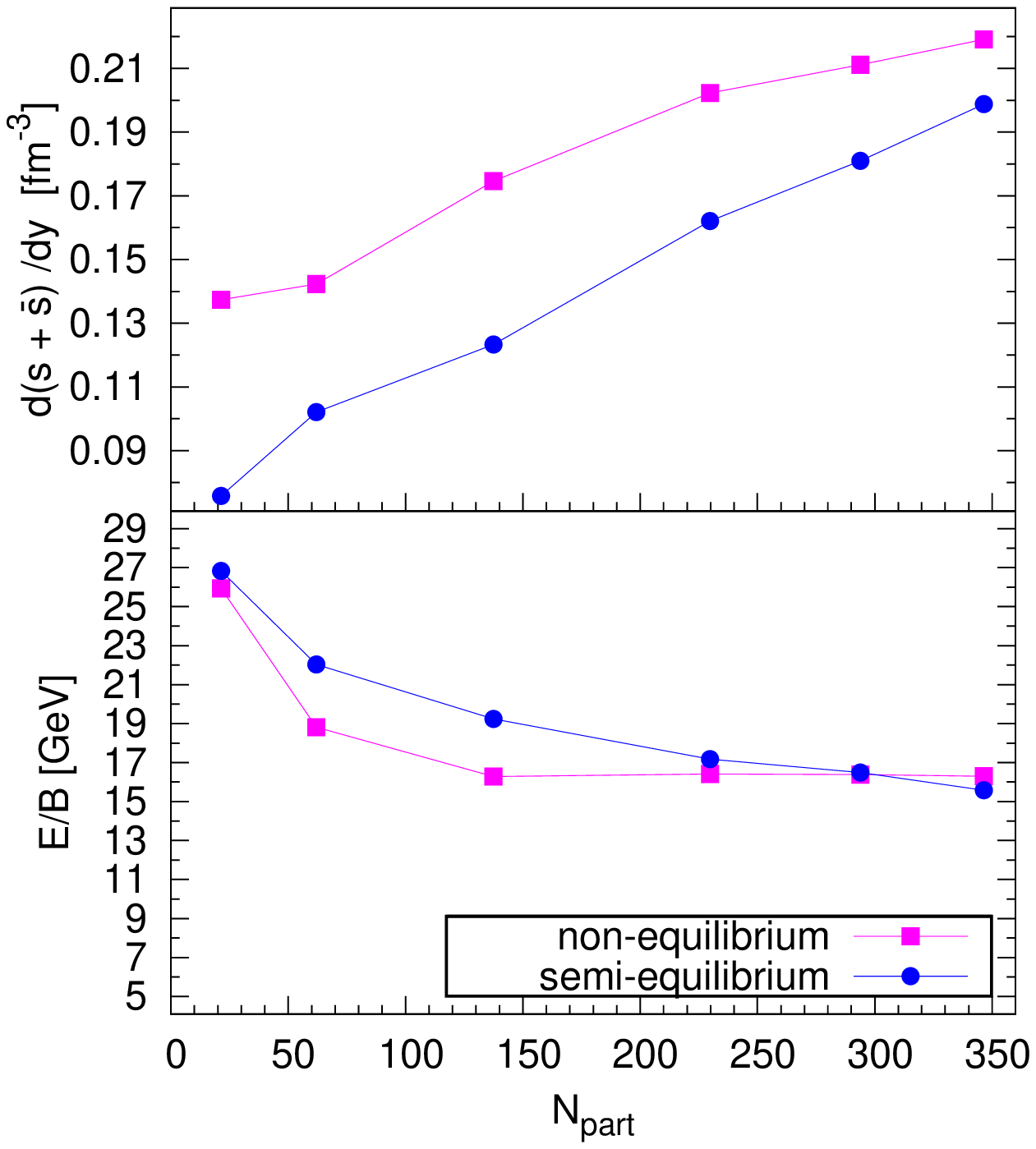}
\end{minipage}
\caption{\label{fig:phys1} {\bf Left panel:} Going from top to bottom, centrality dependence of entropy density energy density ($\sigma$ and $\varepsilon$ respectively) and strangeness per entropy ratio $s/S$.
{\bf Right panel:} Going from top to bottom, centrality dependence of the strangeness rapidity density and energy per baryon.}
\end{figure}

\section{Hadronization conditions}
In the non-equilibrium SHM approach there appears to be universal hadronization condition as function of centrality, which can be seen in figure~\ref{fig:phys1} top two panels on left. Apart from the constant pressure $P \simeq 82\MeV/{\rm fm}^3$, obtained at other RHIC energies~\cite{Rafelski:2009jr}, which value was used in non-equilibrium fits and found consistent with the new data, the entropy density at hadronization stays at constant universal value  $\sigma \simeq 3.3\ {\rm fm}^{-3}$. Similarly, the energy density is constant at $\varepsilon\simeq 0.48 \GeV/{\rm fm}^3$. A universal value of statistical properties at hadronization as function of centrality~\cite{Rafelski:2004dp} supports the notion of a new phase of matter, the quark--gluon plasma undergoing a sudden break up transition to hadrons which free stream, as the chemical non-equilibrium model assumes. On the other hand, in  the semi-equilibrium hadronization model, both entropy and energy density grow nearly proportional to the number of participants.  There is no  physical understanding of this behavior.

We see, in the bottom panel on left in figure~\ref{fig:phys1}, that in SHM non-equilibrium fit, the ratio of strangeness $s$ to entropy $S$ approaches for most central collisions the QGP expectation $s/S=0.031$. The relative yield of strangeness is systematically smaller in the SHM semi-equilibrium fit. On right in figure~\ref{fig:phys1}, we see at the top that strangeness density increases with centrality, in agreement with the expectation that the largest system can approach the QGP yield. Moreover, we see that in the SHM non-equilibrium model, we always find a much greater strangeness yield compared to the SHM semi-equilibrium. The effect is quite large at small centrality, where the entropy contents differ most, see the left top panel. 

Another interesting result in non-equilibrium SHM is the constancy for $N_{\rm part}>130$ of hadronization energy per baryon $E/B$ seen in bottom right of figure~\ref{fig:phys1}. Moreover,
we find that the fraction of the available energy in the collision is universal among other systems and collision energies. We find, $E/B_{\rm most\ central} = 16.5\GeV \simeq 0.25 \times 62.4 \GeV = \frac{1}{4} \sqrt{s_{NN}}$. This result is the same as at the lower SPS energies. It says that the stopping of baryon number is 4 times more effective than the stopping of energy.
Only at small centralities the stopping of energy and baryon number seems to converge to a common value resulting in  $E/B_{\rm most\ peripheral}\to  \frac{1}{2} \sqrt{s_{NN}}$.

\section{Conclusions}
We revisited the statistical hadronization fit of particle yields from Au--Au collisions at $\sqrt{s_{NN}}=62.4\GeV$ recently updated with the yields of multi-strange hadrons. We obtained qualitatively different fit from the one published by the STAR collaboration~\cite{Aggarwal:2010ig} which was done with the constraints $\gamma_q= 1$ and $\gamma_s < 1$ and  which fails to resolve yields of all multistrange particles. Our results show a very good confidence level at all centralities. Our results show that the hadron strangeness phase space is overpopulated  $\gamma_s>1$ from semi-central $N_{\rm part}=150$ to most central collisions $N_{\rm part}=350$. We allowed the light quark phase space, $\gamma_q\neq 1$, to be out of equilibrium as well and the new fit to the experimental data works for multistrange hadrons and has much smaller errors. However, given that $K^0_s$ 
were inconsistent with the yields of charged hadrons, we did fit  $\Lambda+\overline{\Lambda}+2 K^0_s$. We found that the non-equilibrium model provides physical  consistency  in that as a function of centrality, i.e., system size we find appropriate and universal behavior of bulk physical properties: entropy, energy, strangeness and the stopped energy per baryon~\cite{Rafelski:2009jr}.

\section*{Acknowledgments}
Laboratoire de Physique Th{\' e}orique et Hautes Energies, LPTHE, at University Paris
6, is supported by CNRS as Unit{\' e} Mixte de Recherche, UMR7589. This work was 
supported by the grant LC07048 and LA08015 from Czech Ministry of Education and the grant from
the U.S. Department of Energy, DE-FG02-04ER41318.


\begin{thebibliography}{99}%

\bibitem{Torrieri:2004zz}
  G.~Torrieri, S.~Steinke, W.~Broniowski, W.~Florkowski, J.~Letessier and J.~Rafelski,
  Comput.\ Phys.\ Commun.\  {\bf 167}, 229 (2005).
  G.~Torrieri, S.~Jeon, J.~Letessier and J.~Rafelski,
  Comput.\ Phys.\ Commun.\  {\bf 175}, 635 (2006).

\bibitem{Becattini:2010sk}
  F.\,Becattini, P.\,Castorina, A.\,Milov and H.\,Satz,
  Eur.\ Phys.\ J.\  C{\bf 66}, 377 (2010).

\bibitem{Aggarwal:2010ig}
  M.~M.~Aggarwal {\it et al.}  [STAR],
  Phys.\ Rev.\  C {\bf 83} (2011) 024901.

\bibitem{:2008ez}
  B.~I.~Abelev {\it et al.}  [STAR],
  Phys.\ Rev.\  C {\bf 79}, 034909 (2009).

\bibitem{:2008fd}
  B.~I.~Abelev {\it et al.} [STAR],
  Phys.\ Rev.\  {\bf C79}, 064903 (2009).

\bibitem{Petran:2009dc}
  M.~Petran and J.~Rafelski,
  Phys.\ Rev.\  C {\bf 82}, 011901 (2010).

\bibitem{Rafelski:2009jr}
  J.~Rafelski, J.~Letessier,
  J.\ Phys.\ G {\bf G36}, 064017 (2009).

\bibitem{Rafelski:2004dp}
  J.~Rafelski, J.~Letessier, G.~Torrieri,
  Phys.\ Rev.\  {\bf C72}, 024905 (2005).


\end{thebibliography}
\end{document}